# Quantification of the strong, phonon-induced Urbach tails in β-Ga$_2$O$_3$ and their implications on electrical breakdown

Ariful Islam[1], Nathan David Rock[2], Michael A. Scarpulla[1,2]

[1.] Electrical and Computer Engineering, University of Utah, Salt Lake City, USA

[2.] Materials Science and Engineering, University of Utah, Salt Lake City, USA

**Abstract**

In ultrawide bandgap (UWBG) nitride and oxide semiconductors, increased bandgap ($E_g$) correlates with greater ionicity and strong electron-phonon coupling. This limits mobility through polar optical phonon scattering, localizes carriers via polarons and self-trapping, broadens optical transitions via dynamic disorder, and modifies the breakdown field. Herein, we use polarized optical transmission spectroscopy from 77-633 K to investigate the Urbach energy ($E_u$) for many orientations of Fe- and Sn-doped β-Ga$_2$O$_3$ bulk crystals. We find $E_u$ values ranging from 60-140 meV at 293 K and static (structural defects plus zero-point phonons) disorder contributes more to $E_u$ than dynamic (finite temperature phonon-induced) disorder. This is evidenced by lack of systematic $E_u$ anisotropy, and $E_u$ correlating more with X-ray diffraction rocking-curve broadening than with Sn-doping. The lowest measured $E_u$ are ~10x larger than for traditional semiconductors, pointing out that band tail effects need to be carefully considered in these materials for high field electronics. We demonstrate that, because optical transmission through thick samples is sensitive to sub-gap absorption, the commonly-used Tauc extraction of bandgap from transmission through Ga$_2$O$_3$ > 1-3 μm thick is subject to errors. Combining our $E_u(T)$ from Fe-doped samples with $E_g(T)$ from ellipsometry, we extract a measure of an effective electron-phonon coupling indicating increases in weighted 2$^{nd}$ order deformation potential with temperature and a larger value for E||b than E||c. The large electron-phonon coupling in β-Ga$_2$O$_3$ suggests that theories of electrical breakdown for traditional semiconductors need expansion to account not just for lower scattering time but also for impact ionization thresholds fluctuating in both time and space.

**Keywords: ultrawide bandgap semiconductors, Ga$_2$O$_3$, Urbach tails, optical spectroscopy, electrical breakdown**

**Introduction**

Gallium oxide ($Ga_2O_3$) is a promising material for next-generation high-power electronics due to its ultrawide wide bandgap (UWBG), expected high breakdown field, and ability to be grown from the melt as well as epitaxially using multiple processes. $Ga_2O_3$ has many polytypes; among them, β-$Ga_2O_3$ is thermodynamically stable at ambient conditions and has bandgap ($E_g$) estimated to be near 4.8-4.9 eV. Its high melting point near 1800 °C, and strong bonding suggests β-$Ga_2O_3$ for power electronic devices in extreme conditions including e.g. high voltages, high temperatures, and or radiation[1]. High breakdown field ($E_{cr}$) enables higher-efficiency power devices through Baliga's Figure of Merit which scales as $E_{cr}^2$ or $E_{cr}^3$ depending on vertical or lateral device geometry. $E_{cr}$ scales empirically as $\sim E_g^2$ for conventional semiconductors, which led to the suggestion that $E_{cr}$ for $Ga_2O_3$ may be near 8 MV/cm or ~2 times higher than conventional WBG materials like GaN, SiC, or ZnO with Eg=3.2-3.4 eV [2–4]. Detailed calculations taking into account phonon and electronic band structures of β-$Ga_2O_3$ predict anisotropic $E_{cr}$ 5.5-7 MV/cm or 5.8 MV/cm [5–7]. In terms of electron-phonon coupling, these papers focused primarily on carrier scattering (especially polar optical phonon scattering), which acts to prevent charge carriers from accelerating to energies sufficient for impact ionization. However, they did not consider the possible effects of phonons on the ionization threshold energy itself. The first effect of strong electron-phonon coupling clearly would tend to increase $E_{cr}$, while the second is not easy to predict although we speculate that avalanche may be dominated by local fluctuations to smaller threshold energy. We suggest that the effect of ionization threshold fluctuating in space and time is worthy of further investigation.

It is well-known that the optical absorption transitions in monoclinic β-$Ga_2O_3$ are anisotropic; the threshold energies for carrier generation using linearly polarized light depend on crystallographic direction. In a perfect crystal, the onset of a direct optical transition with photon energy would be quite sharp. Disorder, any departure from perfect periodicity, causes a distribution of bandgaps; a missing, extra, or displaced atom at any point changes the local bandgap. Static disorder causes a time-independent, spatially varying bandgap distribution while atoms' random thermal motion due to phonons changes atom positions and thus bandgap dynamically in space and time. The same electron-phonon coupling responsible for the strong polar optical phonon scattering that limits carrier mobility [8–10] mediates this variation in bandgap.

Because optical transitions occur much faster than phonon vibrations, each photon absorption is like a snapshot in time of the dynamic disorder.

This phenomenon of phonon-induced exponential bandgap distributions was first observed by Urbach in silver halides and Martienssen in alkalides[11,12]. The static component of disorder is caused by distributions in bond geometry (non-uniform strain) and any non-periodicity of the crystal from random alloying, point defects, extended structural defects, and in the extreme case, amorphous structure [13]. The nature and concentration of imperfections directly influence the slope and magnitude of the Urbach tails[14,15]. In contrast, dynamic disorder is temperature-dependent and increases with phonon populations as temperature increases. The electron-phonon interaction is the origin of this phenomenon, although the exact details of coupling to each phonon mode are rather complex[16–19].

The basic concept of Urbach tails is familiar; that the joint density of states (and thus $\alpha$) near the bandgap increases exponentially with photon energy hν rather than e.g. $\sqrt{h\nu - E_g}$ as expected for isotropic 3D bands. At fixed temperature, Urbach tails follow $\alpha(E) \propto \exp(\frac{h\nu - E_g}{E_u})$, but this does not take into account the temperature dependencies of $E_g$ or $E_u$. Urbach's rule actually goes further; like the Meyer-Neldel rule for Arrhenius processes [20,21] that relates prefactors to activation energies, Urbach's rule states that that linear fits of ln($\alpha$) vs. hν for different temperatures all intersect in one focus point ($E_o$, $\alpha_o$):[22]

$$\alpha = \alpha_0 \exp \frac{\sigma(T)\,(h\nu - E_0)}{k_B T} \tag{1}$$

In the above, $\alpha$ is the absorption coefficient which describes how much light of a given color will be absorbed by a material in a given thickness, $E_u$ is the Urbach energy (which is distinct from $E_g$), σ(T) is the dimensionless temperature dependent steepness parameter, and $E_u(T) = \frac{k_B T}{\sigma(T)}$. The underlying physics embedded in Urbach's rule is that $E_g(T)$ and $E_u(T)$ change together; for most semiconductors $E_g$ decreases with temperature while $E_u$ increases in such a way to maintain a single focus point. The covariance of $E_u(T)$ and $E_g(T)$ is rooted in the fact that both are related to phonons through the electron-phonon interaction, but the dependence is not trivial [23]. One important point to bear in mind is that measured $E_u$ values result from the combination of disorder in the valence and in the conduction band and it is expected that $E_{u,tot}^2 = E_{u,VB}^2 + E_{u,CB}^2$. From our measurements, we cannot sub-divide these contributions, however, when a single phonon mode of energy $\hbar\omega$ dominates $E_u(T)$, the steepness parameter σ in equation (2) follows the relation[24]

$$\sigma(T) = \sigma_0 \frac{2k_BT}{\hbar\omega} \tanh\left(\frac{\hbar\omega}{2k_BT}\right) \quad (2)$$

In which $\sigma_0$ is a constant independent of temperature. The dominant modes may be different in different materials; for example transverse optical (TO) in CdSe and transverse acoustic (TA) in CdS [24]. In the current case of the anisotropic, multimodal phonon structure in monoclinic β-$Ga_2O_3$, presumably multiple modes will contribute for each incident light polarization, but to our knowledge this has not been computed to date. From optical measurements on β-$Ga_2O_3$, only an effective or averaged mode would be extractable from $E_u(T)$. The model leading to Equation (2) takes into consideration linear interactions between excitons and phonons and assumes a simplified, dispersion less optical phonon mode [23]. It is noteworthy that $\sigma_0$ is inversely related to the the electron-phonon coupling strength; since this is strong in β-$Ga_2O_3$, we expect $\sigma_0$ to be rather small. The temperature dependence of $E_g$ is usually explained as arising from anharmonicity of the lattice leading to thermal expansion which in turn reduces wavefunction overlap as the unit cell expands. However in β-$Ga_2O_3$ it has been shown that the electron phonon coupling is strong enough that displacements within the unit cell caused by temperature-dependent superpositions of phonons also contributes significantly [25].

**Experimental**

We measured optical transmission across a range of Fe-doped and Sn-doped β-$Ga_2O_3$ samples. Sn-doped samples >500 mm thick (data was normalized to the actual thickness for each sample) grown using edge-fed growth (EFG) were obtained from Novel Crystal Technologies (NCT) oriented in the (001), (010), and (-201) directions. These were polished by on both sides by NCT except for the (010) for which we polished the back side to optical flatness. A (100)-oriented piece of a Fe-doped crystal grown by Synoptics using Czochralski (CZ) pulling along the (010) direction was meticulously cleaved using razor blades to a final thickness of 109 μm and 330 μm. All visible thickness steps were removed from both surfaces over an area larger than the measurement beam size and the 109 μm sample used for temperature dependent measurements. Additionally, we measured a 36 mm long, 6 mm diameter rod cored from the same crystal and optically-polished on both of its (010) ends. We also obtained a (010) unintentionally-doped (UID) wafer from Synoptics and polished both sides using a 240-grit SiC pad followed by $Al_2O_3$ grits down to 0.3 µm then final CMP using 0.050 µm colloidal silica. We measured the Fe-doped (100) β-$Ga_2O_3$ samples from 240 to 500 nm using a Hitachi U-4100 spectrophotometer equipped with an integrating sphere detector and a Rochon prism polarizer. We built a custom, short bore 1" diameter furnace and the actual temperature of the sample using a thermocouple in contact with the sample before commencing measurements. To ensure accuracy, we averaged multiple measurements for high signal-to-noise ratio at low transmitted intensity. Sn-doped and UID β-$Ga_2O_3$ samples were measured from 240 to 800 nm in a Perkin Elmer Lambda 900 spectrophotometer with broadband polarizer and integrating sphere detector. Briefly, we determined α assuming insignificant contribution of coherent reflections because of the large thickness compared to wavelengths as $\alpha = -\frac{1}{d}\ln\left(\frac{T}{1-R}\right)$ in which T is the measured transmission and R is the total reflection. Extensive details of the data processing used to convert transmission data to absorption coefficient are given in the supplementary materials.

**Results and Discussion**

Figure 1 presents the room temperature linear absorption coefficient (α) for E||b and E||c determined from a d=330 µm thick sample cleaved on (100) faces from a Fe-doped boule grown by Synoptics. For comparison, we also show the absorption coefficient derived from the imaginary part of the dielectric tensor element for E||b using ellipsometry in [26] ($\varepsilon_{2,zz}$ in the original notation). The fundamental and excitonic critical point energies ($CP^b_{0x}$= 5.4±0.6 and $CP^b_0$=5.6±0.4 eV) from Ref. 26 are indicated. It is immediately clear that the two experiments have measured very different regions of the optical absorption in both energy and α. By measuring many wafer-thick samples and incident polarizations in transmission, we have determined that very strong Urbach tails are present in β-$Ga_2O_3$, manifesting as linear regions on plots of log(α) vs. photon energy. Through careful analysis of uncertainty (see Supplemental Materials), we determined that, although our data shows two apparent slopes, the data for approximately α<10 cm$^{-1}$ is significantly impacted by the precision and accuracy of both measurement and data processing (especially how it is corrected for reflectivity). As discussed below, we only place trust in and further analyze the steeper linear sections of the data for samples herein spanning approximately α>10 cm$^{-1}$ (depending on sample thickness). We note that the Urbach tail for our E||b data extrapolates close

to the CP values from Ref. 26, however measurements were made on different samples thus we treat this as suggestive rather than conclusive.

There is by now widespread understanding that the dielectric or refractive index tensors for monoclinic β-$Ga_2O_3$ contain irreducible off-diagonal terms in the a-c plane containing the non-90° angle, while the b axis is decoupled. Because these ellipsometry experiments were done in reflection, they are capable of measuring both below and above bandgap thus precisely determining critical point energies. However, because the reflected amplitude depends primarily on the real part of the dielectric constant, ellipsometry in this geometry cannot simultaneously accurately measure weak absorption from band tails far below gap. On the other hand, transmission measurements like those herein can accurately measure light intensity from 100% to the noise floor ($I_{noise}$). Fundamentally, no matter what the origin of absorption, samples will extinguish light to intensities below the noise floor for all energies satisfying $I_{noise} = I_o(1-R)\exp(-\alpha d)$ in which all quantities besides d are understood to be energy dependent, $I_o$ is the incident intensity, and R is the total reflection. Determination of the energy-dependent R either by measurement or modelling adds uncertainty. Using Beer's law for a generous case of $I_{noise}/I_o$=0.001-0.01 (2-3 orders of magnitude measurable dynamic range), α can be measured in the range approximately 100-$10^5$ cm$^{-1}$ for 500 nm layers (which would typically be mechanically supported on a wider-gap substrate) and in the range 0.1-100 cm$^{-1}$ for 500 μm thick wafers.

Tauc plot analysis to determine bandgap [27] relies on measuring and extrapolating data for energies >$E_g$ corresponding to absorption coefficients $\alpha \gtrsim 10^4\ cm^{-1}$ for a direct gap transition. This means that Tauc analysis of absorption data from samples possibly having band tails with d greater than a few μm should be suspect because optical thickness αt<5 may occur in the band tail region at energies significantly below $E_g$. This can cause significant underestimation of $E_g$ if this fact is not recognized. In Fig. 1b) we illustrate an extreme case of this effect for β-$Ga_2O_3$ by applying Tauc analysis to data we deem reliable covering α=0.1-1 cm$^{-1}$ from a 36 mm long (010) Fe-doped rod cored from the same Synoptics boule that supplied the Fe-doped sample in Fig. 1a) and having polished ends. Despite the continuous curvature of the data in Fig. 1b), if we apply the typical process of linear extrapolation from the highest energy data, this suggests a "bandgap" of 3.1-3.3 eV which is obviously in error. The polarized photocurrent onset of β-$Ga_2O_3$ is around 4.6 eV depending on the orientation of the crystal[28] and applied bias. Again, this is a result of applying

Tauc analysis to a set of inappropriate absorption data not covering the range >$E_g$ because for this very thick sample the band tails extinguish the intensity for energies well below $E_g$. We speculate that some of the uncertainty and ranges of values for optical absorption thresholds previously-published for β-$Ga_2O_3$ may arise from these underlying experimental issues.

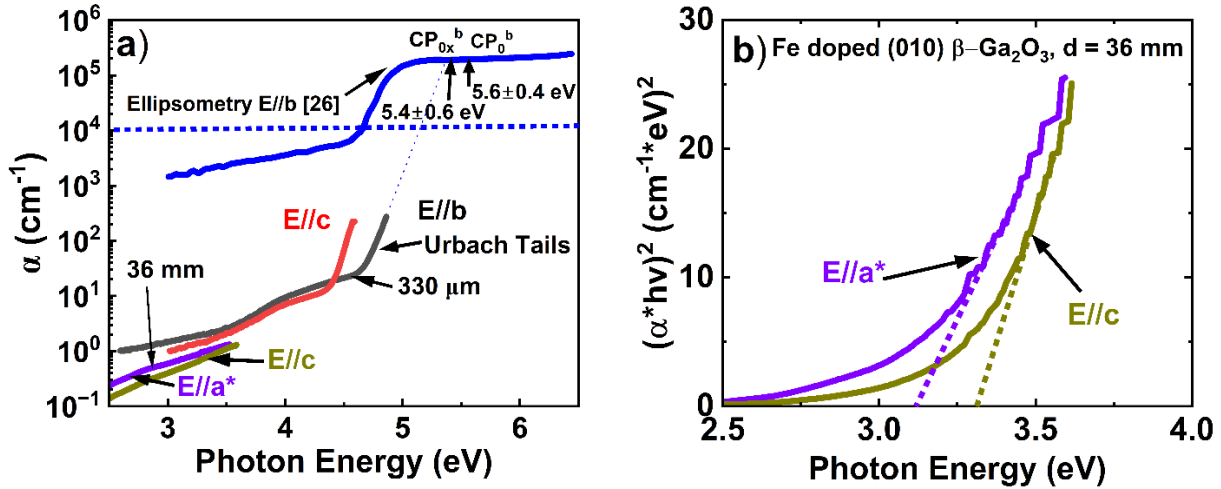

Figure 1 – a) Room temperature β-$Ga_2O_3$ absorption coefficients determined from spectroscopic ellipsometry [26] and from optical transmission measurements herein for a 36 mm rod with (010) polished faces and 330 mm thick (100) sample cleaved from the same Fe-doped boule. Note the vastly different ranges of a measurable for different techniques and sample thicknesses. The grey dashed line shows the extrapolation of our Urbach tail to Mock's data, suggesting approximate agreement. b) To illustrate the erroneous bandgap estimates possible from data sets not actually measuring above-gap absorption, Tauc analysis was carried out on the data from the 36 mm rod. Because the data satisfies at>3-5 from sub-gap transitions, it suggests E||a* and E||c "bandgaps" of 3.1-3.3 eV, which are at least 1 eV below the generally-reported values for b-$Ga_2O_3$. We thank Prof. A. Mock for sharing the numerical values of the dielectric tensor from Ref.[26] which we converted to absorption coefficient for this work.

As discussed in the Supplementary Materials, we carefully examined the effects of data processing, especially the numerical subtraction of reflectivity given by an index of refraction assumed constant or as a Cauchy function on the extracted values of α. The regions appearing as straight lines in Fig. 1a) for α>10-20 cm$^{-1}$ were insignificantly affected by various reflectivity subtraction methods and thus further analysis is done exclusively on these regions of data sets to extract $E_u$. The reasonable matching of data in Fig. 1a) from the 36 mm rod with that from the cleaved wafer gives some confidence that the deep sub-gap region has been determined with fair accuracy although the details of reflectivity subtraction do change the detailed shape and magnitude of the extracted a below the Urbach tail region.  Also, it could be seen by eye that the Fe distribution was not completely homogeneous through the boule grown by Synoptics in one of their first growth campaigns.  Although we refrain from detailed analysis we speculate this deep sub-gap region reflects absorption by impurity (e.g. Fe) and native defect states as internal excitations and or charge transfer excitations measurable by photocapacitance methods such as DLOS [29].

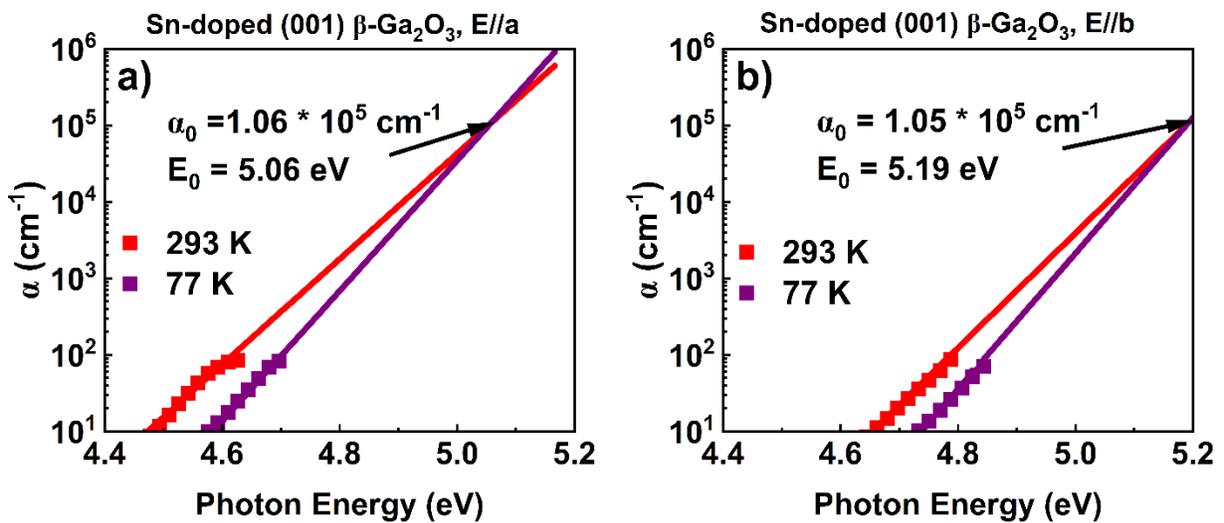

Figure 2 - α for Sn-doped (001) $\beta$-$Ga_2O_3$ for a) E//a and b) E||b at 293K and 77K temperatures.  Data is shown as points, while lines result from global fitting using Eq. 1 with $\alpha_o$ and $E_o$ constant for both temperatures but different $\sigma_o$ values for each temperature.

In this investigation, we endeavored to investigate the contributions to $E_u$ from structural defects and phonons, as well as their anisotropies.  We note that phonon modes are excited even at 0 K by zero-point energy, as well as the thermal population which increases with temperature. First, we attempted to separate the temperature-independent (zero-point phonon + structural defects) from temperature-dependent (thermally-populated phonon) disorder by cooling to cryogenic temperatures.  Figure 2 shows the Urbach tail region of α for a Sn (001) $\beta$-$Ga_2O_3$ wafer at room and liquid nitrogen temperature.  This sample exhibited the smallest values of $E_u$ we measured – still ~ 60 meV at room temperature - but cooling to 77 K only reduced $E_u$ by approximately 10 meV.  The characteristic phonon energies in $\beta$-$Ga_2O_3$ range from approximately 19-93 meV [30], thus the phonon populations (which should be proportional to phonon-induced disorder via the mean-square displacement $u^2$) for the various modes should be suppressed by factors ranging from ~15 to 25,000 at 77 K. Cooling closer to 0 K would be expected to further reduce the measured $E_u$, although perhaps not dramatically as discussed below. This suggests that at room temperature and below, the majority contribution to $E_u$ in these bulk crystal samples arises from temperature-independent zero-point phonons and structural defects, rather than temperature-dependent disorder from thermal occupation of phonon modes. The ratio of temperature-independent to temperature-dependent contributions to $E_u$ appears to be near 5:1.

Similar data was measured at room temperature for a wide variety of Sn-doped samples, Fe doped (100) and a UID (010) sample for different crystal orientations and polarizations. The data are summarized in Fig. 3 along with those shown in Figs. 2. Since two orthogonal in-plane incident light polarizations were investigated for each sample, Figure 3 shows each pair of measurements from the same sample as two points connected by a line as a guide to the eye. Clearly $E_u$ varies quite radically with orientation within each sample, with doping, and to some degree with temperature.

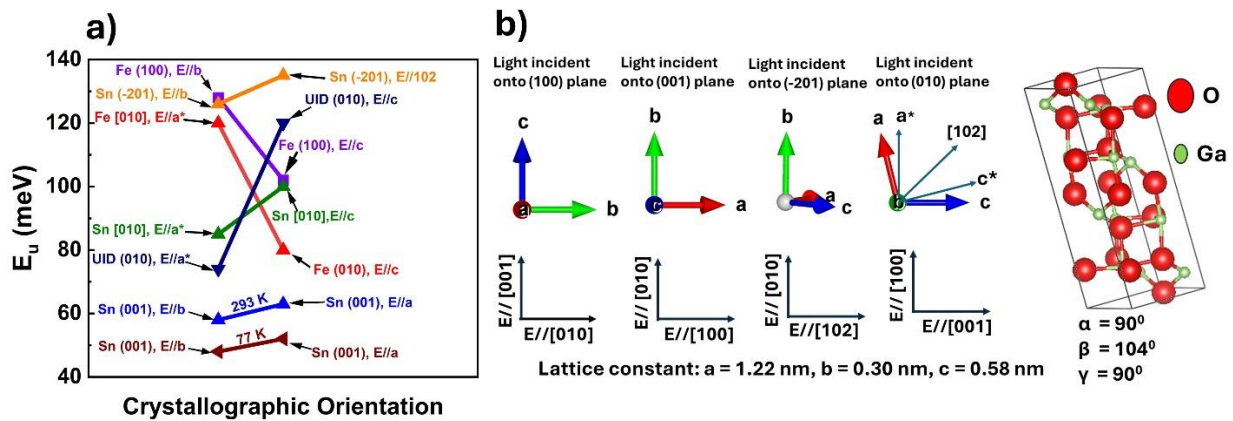

Figure 3 – a) Summary of measured Urbach energies for Fe-doped, Sn-doped, and UID β-Ga$_2$O$_3$ wafers of different crystallographic orientations, labeled by the plane of the wafer. For each different sample and temperature, two values connected by a solid line are shown to help highlight the in-plane anisotropy for orthogonal incident light polarizations. All data are at 293 K except the one labeled as 77 K. b) Schematic of sample orientation, axes, polarization and atomic unit cell of β-Ga2O3. The lattice constant parameters is taken from[31].

First, we note that across all of the samples we measured, $E_u$ was ~60-140 meV at room temperature which is a very large value compared to those typical for conventional single crystal semiconductors; typically values of a few meV are expected for single crystalline, non-alloyed semiconductors like epitaxial III-V and II-VI semiconductors while values of a few 10's of meV might be observed for high-quality randomly alloyed semiconductors [32]. Thus, at least in these bulk crystals we have measured, the $E_u$ values are quite significantly higher than those typical for more familiar semiconductor materials.

Next, we examined the possibility of an underlying crystalline anisotropy in $E_u$ across all the investigated samples; unfortunately, we cannot deduce any universal trends although we can not rule out the possibility of such effects in more perfect samples. For some samples $E_u$ is larger for E||a, others for E||b, etc. Thus, from this data set using melt-grown bulk crystals, we deduce that the extrinsic disorder from specific sets of defects present in different samples dominates $E_u$ rather than intrinsic crystalline anisotropy. Samples with much higher crystalline perfection would need to be measured in order to accurately measure the intrinsic crystalline anisotropy in $E_u$. We note that at the small absorption coefficients associated with Urbach tails, surface layers of a few μm of highly-damaged material e.g. from cutting, lapping and polishing could significantly contribute to the total sub-gap absorption.

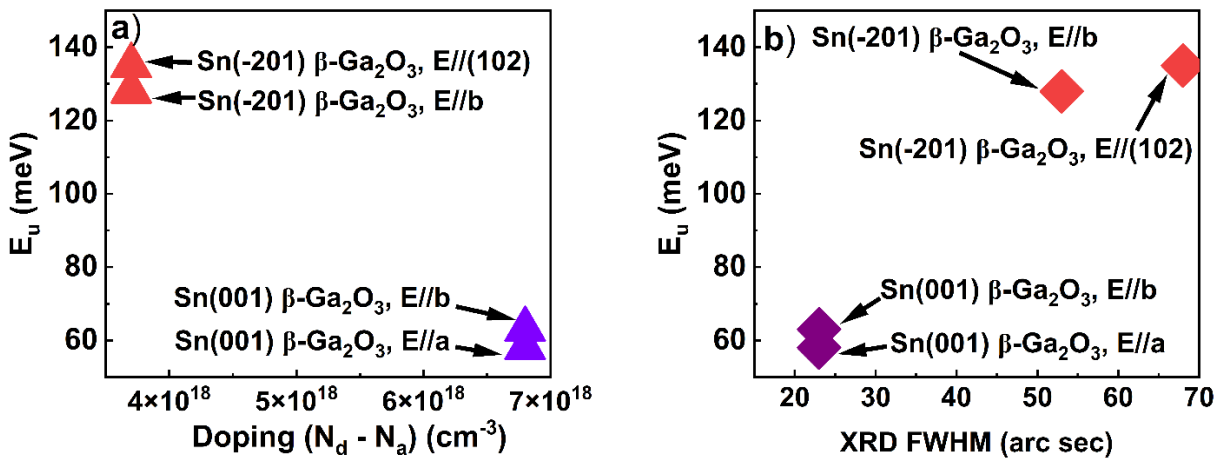

Figure 4 – a) $E_u$ vs. [Sn] and b) $E_u$ vs. on-axis x-ray diffraction FWHM reported by NCT for these wafers. $E_u$ would be expected to increase with Sn doping based on non-uniform strain arguments as well as the expected increasing number of compensating native defects (e.g. $V_{Ga}$ and related complexes). The fact that this is not observed suggests that other sources of structural disorder to which the x-ray diffraction experiments are sensitive dominate $E_u$.

In order to gain insight into the types of disorder primarily responsible for $E_u$, we examined the set of Sn-doped wafers of various orientations obtained from NCT since they are all produced from boules grown along the b-axis using EFG but cut on different crystal planes. We hypothesized that that the degree of non-uniform strain and numbers of induced compensating native defects would scale with the concentration of Sn doping (which differed within the $10^{18}$ /cm$^3$ range for the samples we measured), and thus $E_u$ should scale with this variable. However, we found that the $E_u$ actually decreased with the concentration of Sn as shown in Fig. 4a). For each wafer we also had the on-axis Xray rocking curve FWHM specified by NCT. With the exception of one data point not shown, Fig. 4b) indicates better correlation between $E_u$ and the measured FWHM. On-axis Xray rocking curves measure crystalline disorder in the same plane measured by incident polarized light (the width of θ-2θ scans measures disorder in the depth direction). The Xray rocking curves will probe both the doping- and point defect induced disorder and that arising from extended defects and near-surface polishing damage. Thus, we conclude that later two effects are dominant contributions to $E_u$ in these samples. From our data we can not discern whether bulk structural defects or near-surface damage from wafer preparation (which can also be anisotropic) is the primary contributor.

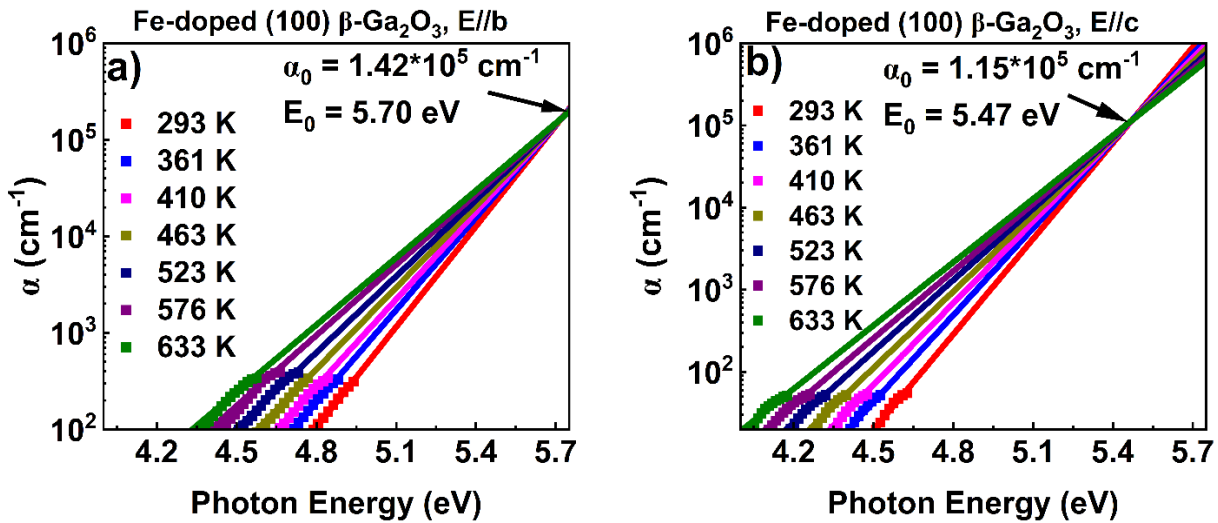

Figure 5 - Extracted α(E,T) for Fe-doped (100) β-Ga$_2$O$_3$ as a function of photon energy for a) E||b and b) E||c at varying temperatures up to 633 K. Measured data is shown as points, while the lines show the

results of global fitting to Eq. 1 with $\alpha_o$ and $E_o$ constant but $\sigma_o$ different for each temperature. From these data taken from the same sample we can extract information about the magnitude and anisotropy of the phonon-induced contributions to $E_u$.

Since many samples were measured under many conditions, we introduce Table I to assist the reader.

Table-I: Summary of samples measured with different incident polarizations and temperatures. Despite our knowledge that Urbach tails rather than above-gap absorption were measured and thus our data is inappropriate for Tauc analysis, the rightmost column gives the "bandgap" that would be obtained from our absorption data if it were blindly carried out. Please note that we can not determine the actual bandgap from our data because sufficiently thin samples were not available.

| Doping | Wafer Orientation | Polarization of incident light | Thickness of the sample (μm) | Temperature (K) | Bandgap from Tauc analysis (eV)* |
|---|---|---|---|---|---|
| Fe-doped | (100) | E//b | 109 | 293 - 633 | 4.78 - 4.42 |
| Fe-doped | (100) | E//c | 109 | 293 - 633 | 4.51 - 4.03 |
| Fe-doped | (100) | E//b | 330 | 293 | 4.74 |
| Fe-doped | (100) | E//c | 330 | 293 | 4.45 |
| Fe-doped | (010) | E//c | 695 | 293 | 4.45 |
| Fe-doped | (010) | E//a* | 695 | 293 | 4.42 |
| Fe-doped | (010) | E//c | $36 \times 10^3$ | 293 | 3.3 |
| Fe-doped | (010) | E//a* | $36 \times 10^3$ | 293 | 3.17 |
| Sn-doped | (001) | E//a | 660 | 293, 77 | 4.51, 4.62 |
| Sn-doped | (001) | E//b | 660 | 293, 77 | 4.54, 4.78 |
| Sn-doped | (010) | E//c | 520 | 293 | 4.44 |
| Sn-doped | (010) | E//a* | 520 | 293 | 4.51 |
| Sn-doped | (-201) | E//b | 700 | 293 | 4.65 |
| Sn-doped | (-201) | E//[102] | 700 | 293 | 4.47 |
| UID | (010) | E//c | 673 | 293 | 4.45 |
| UID | (010) | E//a* | 673 | 293 | 4.40 |

We next examine the temperature-dependent contributions to $E_u$ through measurements at elevated temperatures. Figure 5 shows the extracted absorption coefficients for temperature dependent data on a Fe-doped (100) sample for a) E‖b and b) E‖c incidences and global fitting (data from all temperatures fit together) to Eq. 1. Separately for the two incident linear polarizations, the Urbach focus ($\alpha_o$, $E_o$) is determined from all temperature data, while $E_u$ (and thus $\sigma_o$) change with temperature. We estimate uncertainties of ±2 meV for all Urbach energies, arising from both noise in the data and the variation induced in numerical data processing. The values of $E_o$ and $\alpha_o$ are indicated in the figure for each incident polarization, and the extracted $E_u(T)$ values up to 633 K are plotted as Fig. 6. Data obtained at higher temperatures was less reliable and repeatable because of the onset of blackbody emission from the furnace and the more significant convection-driven turbulence in the open-air optical path and thus is not included in further analysis. The onset of significant changes in native defect concentrations in n-type $Ga_2O_3$ is known to be above ~900 °C in air, although it is possible that redistributions of impurities may occur. The fact that these wafers were grown from the melt and cooled through temperature ranges much greater than those used for measurements argues against this being significant.

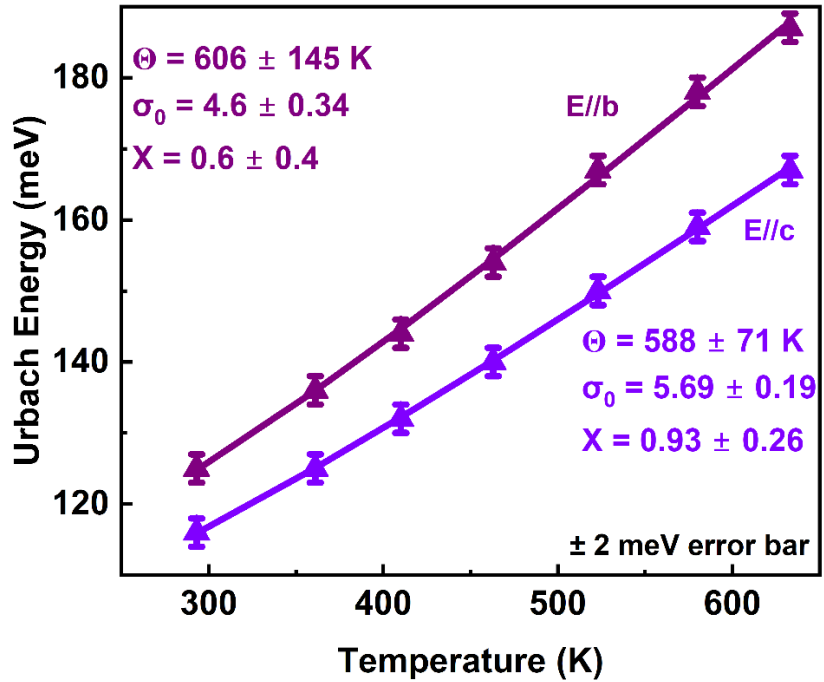

Figure 6: Eu(T) for a E||b and E||c incident polarization directions for Fe-doped (100) β-$Ga_2O_3$ extracted from the data in Figure 2. The solid lines are the best-fits to Eq. 3 and the fitting parameters are noted in the figure.

We analyze the data in Fig. 6 within the framework first introduced by Cody et al., for analysis of a-Si. The premise underlying Cody's decomposition of the measured Urbach energies into structural and phonon-related contributions is that the degree of atomic displacement from their periodic lattice sites caused by either static structural disorder or atomic vibrations is linked to the measured $E_u$. It is only non-uniform strain that causes broadening of the absorption edge, and likewise the first moment of atomic displacements averaged in time caused by phonons is zero. Thus the $E_u = \frac{k_B T}{\sigma}$ should be proportional to the mean squared displacement $\langle u^2 \rangle$ which is the quadrature sum of the static structural, zero-point phonons and thermal components [33–35]. We slightly modify Cody's original formula to highlight the three contributions to total $E_u$:

$$E_u(T,X) = K\left[\langle u^2\rangle_{X,0} + \langle u^2\rangle_{Ph,0} + \langle u^2\rangle_{Ph,T}\right] = \frac{\Theta}{\sigma_0}\left[\frac{X}{2} + \left(\frac{1}{2} + \frac{1}{\exp\left(\frac{\Theta}{T}\right)-1}\right)\right] \quad (3)$$

In the first expression, K is the proportionality constant, $\langle u^2\rangle_{X,0}$ is the temperature-independent contribution of structural disorder, $\langle u^2\rangle_{Ph,0}$ represents the temperature-independent zero-point excitation of phonons and $\langle u^2\rangle_{Ph,T}$ is the temperature-dependent contribution of thermal phonon occupation. In the second expression, K is defined as $\frac{\Theta}{\sigma_0}$, the ½ in the second term represents the zero-point phonon displacements and X characterizes displacements from structural defects such that X=1 would be equivalent to the zero point phonon displacements. The zero-point excitation of the lattice produces a finite Urbach energy at X=0 and T=0, $E_u(0,0) = \frac{\Theta}{2\sigma_0}$. The mean square thermal displacements are proportional to the population of phonons, given by the Einstein model in which $\Theta$ can be recognized as $\frac{\hbar\omega}{k_b}$. Cody points out that the Debye temperature modeled for acoustic modes $\Theta_D = \frac{4}{3}\Theta$. Because of the large number of phonon modes within a small energy range for β-Ga$_2$O$_3$, it is not possible to separate their contributions from temperature-dependent $E_u$ measurements alone. Thus, we take $\Theta$ to represent a properly-averaged representative phonon mode.

Figure 6 shows the results of fitting of our measured $E_u$ data for Fe-doped (100) β-Ga$_2$O$_3$ with Eq. 3. The uncertainty estimates were derived from sensitivity analysis of the fitting (thus the different uncertainties for the two data sets). The best fit parameters are $\Theta$=606 ± 145 K for E‖b and 588 ± 71 K for E‖c corresponding to Debye temperatures of 808 ± 193 K and 784 ± 93 K for E‖b and for E‖c respectively. These values are in reasonable agreement with the experimental value of 738 K[36] and the predicted value from first principles calculations of 872 K[37]. We find the steepness parameter $\sigma_0$=4.60±0.34 (5.69±0.19) and structural disorder X=0.60±0.40 (0.93±0.26) for E‖b (E‖c). Examining the static structural disorder, we note that for E‖b its mean square displacement is about half that of the zero-point phonons, while for E‖c the two effects are nearly equal. Thus, the zero point phonon contributions are larger for both directions than either the structural defect or finite-temperature phonon contributions, despite X being larger for E||c, it is offset by the proportionality constant $\sigma_0$ also being higher which ultimately results in measured $E_u$ values for E||b being larger overall. Using the fitting values and Eq. 3, we determine that the $E_u$ at 0 K are predicted to be 106 meV for E||b and 100 meV for E||c for this Fe-doped sample. For E||b, 69 meV would be from zero-point phonons and 37 meV from structural disorder, while for E||c 52 meV would be from zero-point phonons and 48 meV from structural disorder. Thus, we find evidence for intrinsic anisotropy in static structural $E_u$ of order 17 meV and minimum possible values at room temperature of 88 meV for E||b and 68 meV for E||c. The breakdown of the three different contributions to $E_u$ at different temperatures and for different polarizations are summarized in Table II.

Table II – Analysis of contributions to $E_u$ based on fitting our elevated temperature data from Fig. 6 with Eq. 3. In the rightmost column, we compute the ratio of apparent anisotropy to its uncertainty. If this is >1 the anisotropy is statistically significant, while there is no clear evidence for anisotropy otherwise.

|  | $E_u$ E||b (meV) | $E_u$ E||c (meV) | Anisotropy $E_{u,b} - E_{u,c}$ (meV) | Anisotropy/ Uncertainty Ratio |
|---|---|---|---|---|
| $E_u$ measured 293 K | 125±2 | 116±2 | 9±3 | 3.0 (significant) |
| $E_u$ (X, 293 K) from Eq. 3 | 125±35 | 116±17 | 9±39 | 0.23 (ambiguous) |
| $E_u$ (0,293 K) room temperature contributions of zero point and thermal populations of phonons | 85±11 | 68±4 | 17±12 | 1.4 (somewhat significant) |
| $E_u$(X, 293 K) - $E_u$ (0, 293 K) contribution of static structural disorder | 40±11 | 48±4 | 8±12 | 0.66 (ambiguous) |
| $E_u$( X, 0 K) Total temperature-independent (structural defects plus zero point phonons) | 106±38 | 100±18 | 6±42 | 0.14 (ambiguous) |
| $E_u$(0, 0 K) zero-point phonons only | 66±17 | 52±7 | 14±18 | 0.78 (ambiguous) |
| $E_u$(X, 0 K) - $E_u$(0, 0 K) static structural disorder | 40±29 | 48±15 | -8±33 | 0.24 (ambiguous) |

From Fig. 6 and Table II, we have measured anisotropy in $E_u$ between b and c directions at elevated temperatures. Using the fitting values from Eq. 3, we can deduce the various causes of $E_u$ and its anisotropy. The fitting from the sample used for elevated temperature measurements, with X=0 gives an estimate for the phonon-only minimum possible values of Eu at 293 K (85 and 68 meV) and at 0 K (66 and 52 meV) for b and c directions respectively. As discussed before, these values are 2-10 times larger than values for more traditional semiconductors. We do note that the (001) Sn-doped sample we measured which showed $E_u$(293 K) slightly larger than 60 meV demonstrates that slightly smaller values are possible; we suspect but can not prove from the present data sets that very thin but highly damaged surface layers related to polishing may contribute on the order of ~10 meV, which would bring the uncertainties between samples close to alignment. Looking at the breakdown of contributions to $E_u$ at 0 K, first we define the bottom 3 rows of Table II. $E_u$(X,0) denotes the modeled Eu using T=0 and the fit parameters from Fig. 6, while Eu(0,0) is only the zero-point phonon component and the final row is computed as $\frac{\Theta X}{2\sigma_0}$. We see that anisotropy in the modeled zero-point phonon component $E_u$(0,0) is just below the threshold for being significant, while both the total $E_u$ and the isolated component for static structural defects are isotropic within our uncertainties. Thus, since anisotropy in X (from structural defects) would be the same at 0 K and elevated temperatures, we conclude that the anisotropy in $E_u$ observed at 293 K is caused by phonon anisotropy. At the highest temperature of 633 K, we measure 187 meV and 167 meV for E||b and E||c respectively yielding a 20 meV anisotropy with uncertainty estimated as 2-3 meV. Evaluating Eq. 3 with the best fit parameter values and taking into account the parameter sensitivity of the fitting gives the same values but with much higher uncertainty. Thus, we conclude experimentally that there is some minor anisotropy on the order of 5% of the total $E_u$ at room temperature induced by the anisotropic phonons in β-$Ga_2O_3$.

We next interpret the temperature dependence measured and estimate the electron-phonon coupling implied, which should also affect carrier mobility and $E_{cr}$. Cody et al. [35] also posit that temperature dependence of the energy gap, which arises solely from phonon-induced distortions of the lattice, can be written as

$$E_g(T) = E_g(0) - D\left(\langle u^2 \rangle_T - \langle u^2 \rangle_0\right) \tag{4}$$

in which $E_g(0)$ is the optical bandgap gap at 0 K and $D = \frac{\partial^2 E}{\partial r^2}$ is a second-order deformation potential (eV/Å²), which characterizes the electron-phonon coupling. This deformation potential D specifies the second order derivative of band energy with respect to atomic displacements. This second order deformation potential is different from the first-order deformation potential which is relevant for acoustic phonon scattering. This formulation implies that any bandgap changes caused by static disorder (from both zero-point phonons and structural imperfection) contribute to $E_g(0)$. Herein we take D as a scalar although generalization to tensor form would of course be possible in more advanced treatments. Since the mean-squared displacement increases as phonon mode occupation, we see that this quantity is also related to the derivative of eigenvalue energies with respect to phonon occupation [19]. Since the temperature dependencies of both $E_u$ and $E_g$ share the same root cause of phonons, the mean square lattice displacement in equation (4) can be written as in terms of $E_u$ leading to [35]:

$$E_g(T,X) = E_g(0,0) - \langle u^2 \rangle_0 D \left( \frac{E_u(T,X)}{E_u(0,0)} - 1 \right) \tag{5}$$

Rearranging equation (5) to take the ratio of the two temperature dependencies yields an expression for the quantity $\langle u^2 \rangle_0 D$ (in eV) which would be the magnitude of the electron-phonon coupling at 0 K.

$$\langle u^2 \rangle_0 D = - \frac{E_g(T,X) - E_g(0,0)}{\left( \frac{E_u(T,X)}{E_u(0,0)} - 1 \right)} \tag{6}$$

Thus, we can extract this measure of the electron-phonon coupling by combining temperature dependent measurements of both $E_g$ and $E_u$. As per the discussion of Fig. 1, ellipsometric measurements of $E_g$ are more reliable unless few-mm thick samples are available for transmission, while transmission measurements of $E_u$ are more reliable. First, we observe that small amounts of static structural disorder should have negligible effect on $E_g$; thus, we assume that $E_g(0,X) \approx E_g(0,0)$ – this is also justified by the dominance of zero-point phonons in the static disorder. Making this assumption, the numerator of Eq. 6 is calculated at each of our measured temperatures from the Bose-Einstein parameterization of the $E_g(T,X) - E_g(0,X)$ given in Ref. [26]. $E_u(T,X)$ is our measured temperature-dependent Urbach energy data and $E_u(0,0)$ is given in Table II. The extracted values of $\langle u^2 \rangle_0 D$ are given in Fig. 7 below. We note that Ga$_2$O$_3$ has many modes of different energies thus the quantity we extract represents some weighted average over our measured temperature range including any subtle differences between phonons' effects on $E_u$ and $E_g$. The values we extract consistently increase with temperature, which we interpret as arising from this population-weighting of phonon modes that would favor lower-energy modes at lower temperatures but include all modes at very high temperatures.

The product $\langle u^2 \rangle_0 D$ gives a measure of the electron-phonon coupling as an energy, which we find to have some temperature dependence. Because the zero-point displacements are by definition temperature independent, this must arise from the population weighting of phonon modes which have different effective D values. Phonons in β-$Ga_2O_3$ are plentiful, complex, and anisotropic, with energies spanning approximately 10-100 meV [30]. An interesting open question is how exactly to quantify the coupling between the phonons' displacements and or induced dipoles and the distribution of bandgaps at each temperature that generate the Urbach edges. The simplest but most computationally expensive method would appear to be direct calculation of bandgaps in a large number of supercells with different superpositions of the phonon displacements applied and fitting the Urbach formula to the resulting distribution of bandgaps.

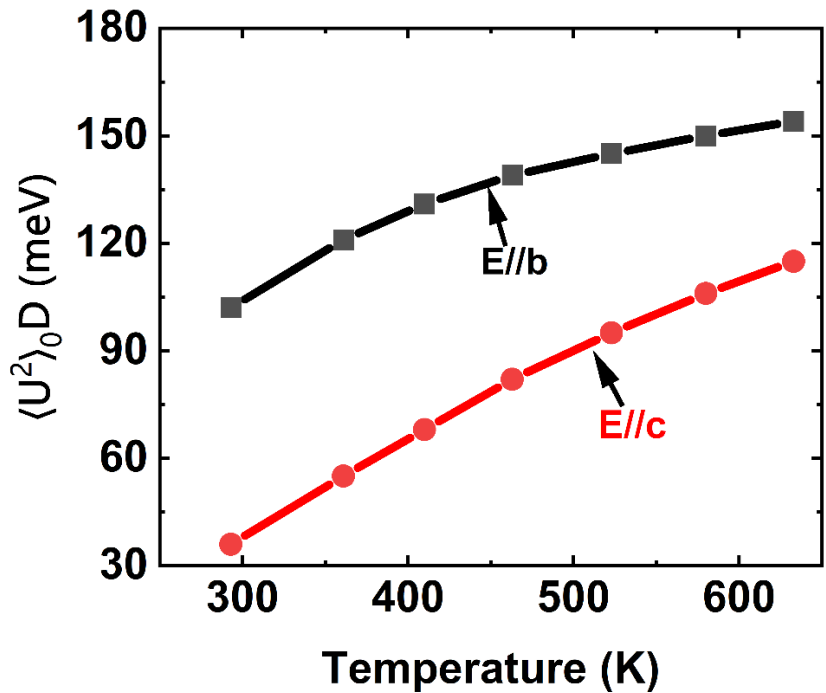

Figure 7: $\langle u^2 \rangle_0 D$ extracted for Fe-doped (100) β-Ga$_2$O$_3$ in E||b and E||c at temperatures from 293 K to 633 K. These represent averages of the electron-phonon coupling across subsets of the full phonon spectrum, properly summed in k-space, and accounting for phonon occupations at each temperature.

It has been established that the dominant intrinsic carrier scattering mechanism (i.e. excluding ionized-impurity scattering) in β-Ga$_2$O$_3$ at room temperature is polar-optical phonon (POP) scattering. In compound semiconductors with polar bonds, Frölich coupling is typically assumed in which carriers scatter from the fluctuating dipole field induced by transverse-optical phonons. The 2$^{nd}$ order deformation potential extracted here is not identical to the Frolich coupling energy responsible for POP phonon scattering [6]. While E$_u$, mobility, and breakdown all depend on electron phonon coupling, they each have different dependencies and must be carefully interpreted.

Despite its seeming straightforwardness, electrical breakdown in inorganic crystalline materials like β-Ga$_2$O$_3$ is a complex topic. The fact that Ga$_2$O$_3$ exhibits strong Urbach tails predicts qualitatively that the mobility and thus mean free path for carrier acceleration in a high field are curtailed, making it less likely for carriers to reach kinetic energies sufficient to impact ionize. However, an unexamined topic (to our knowledge) is how the energy required for impact ionization varies in the presence of significant time-and-space bandgap disorder associated with Urbach tails. For example, if we use the empirical scaling that breakdown field is proportional to E$_g^2$ we would expect variation in E$_{cr}$ with position in Ga$_2$O$_3$. These two physical effects associated with phonon-induced bandgap disorder push the critical breakdown field in different directions thus the net effect is difficult to predict. Breakdown in a-Se and a-Si:H – extreme examples of static structural disorder having Urbach tails on-par with what is observed for Ga$_2$O$_3$ - has been studied and in general it is found that avalanche multiplication in these materials occurs at higher fields than for ordered semiconductors with similar bandgap materials[38]. This argues that in the case of *static structural disorder*, the lower mobility prevents carriers from easily accelerating to high energies in a field. The effects of *phonon-related* disorder – bandgap fluctuations and carrier scattering from phonons especially in the case of large electron-phonon coupling as in β-Ga$_2$O$_3$ – have not been elucidated. If the effects are similar to structural disorder, which may be likely since the root cause of both is mean squared displacement from ideal sites, we may expect that materials with higher electron-phonon coupling might have even higher breakdown fields than their counterparts of the same bandgap. It has been established that POP phonon scattering should reduce the empirically predicted breakdown field in perfectly-crystalline $Ga_2O_3$ of ~8 MV/cm to ~5-6 MV/cm [5], but the effects of the combination of thermal and structural disorder have not been accounted to date. This issue will have a profound effect across all materials because it has not been incorporated into theoretical frameworks and thus has priority for future investigations.

**Conclusions**

We investigated the sub-gap absorption of a number of bulk β-Ga2O3 wafers using linearly-polarized transmission. Overall, we find significant Urbach tails that range from 60 to nearly 140 meV at room temperature depending on doping, crystallographic direction, and individual sample. Static structural defects and zero-point phonons accounts for the majority of the effect, but a non-negligible component arises from finite temperature phonons. There is some evidence for the phonons inducing some anisotropy into the Urbach energies, although this is at the 10% level so not dominant compared to the mean isotropic values. We predict based on comparison to results from amorphous materials, that UWBG materials with high electron-phonon coupling may have even larger breakdown fields than predicted using scaling from conventional semiconductors where this is weaker. Detailed theory will be needed to resolve whether or not the impact ionization threshold also varies as a result of static structural disorder and whether this changes $E_{cr}$.

**Supplementary Material**

The supplementary materials include extensive details of the data processing used to convert transmission data into absorption coefficients, along with the associated error estimates.

**Acknowledgements:**

The authors would like to acknowledge the funding provided by the Air Force Office of Scientific Research under Award FA9550-21-0078 (Program Manager: Dr. Ali Sayir). We thank Alyssa Mock for sharing the ellipsometry data used for Figure 1, and acknowledge fruitful discussions with Profs. Zlatan Aksamija, Elif Ertekin, Feliciano Giustino, and Dragica Vasileska.

# Authors Declaration

## Conflict of Interest

The authors have no conflict of interest.

## Data Availability

The data that support the findings of this study are available within the article. Additional data are available from the corresponding author upon reasonable request.